# Security Policies for WFMS with Rich Business Logic — A Model Suitable for Analysis

Fábio José Muneratti Ortega[1], Wilson Vicente Ruggiero[2]
*Departamento de Computação e Sistemas Digitais*
*Escola Politécnica da Universidade de São Paulo*
São Paulo, Brazil
[1]fortega@larc.usp.br
[2]wilson@larc.usp.br

***Abstract*—This paper introduces a formal metamodel for the specification of security policies for workflows in online service systems designed to be suitable for the modeling and analysis of complex business-related rules as well as traditional access control. A translation of the model into a colored Petri net is shown and an example of policy for an online banking system is described. By writing predicates for querying the resulting state-space of the Petri net, a connection between the formalized model and a higher-level description of the security policy can be made, indicating the feasibility of the intended method for validating such descriptions. Thanks to the independent nature among tasks related to different business services, represented by restrictions in the information flow within the metamodel, the state-space may be fractioned for analysis, avoiding the state-space explosion problem. Related existing models are discussed, pointing the gain in expressiveness of business rules and the analysis of insecure state paths rather than simply insecure states in the proposed model. The successful representation and analysis of the policy from the example combined with reasonings for the general case attest the adequacy of the proposed approach for its intended application.**

## I. INTRODUCTION

In spite of the many advances in security policies description, modeling and validation, designing secure systems under security constraints involving business parameters can lead to large models that are unsuitable for analysis. Additionally, descriptions of security policies based on entities with a high level of abstraction result in models distant from the system's implementation, potentially leading to the inclusion of vulnerabilities in the translation or, if methodically or automatically translated, may still lead to inefficient software.

Our objective is to define a modeling and analysis strategy that best suits the validation of security policies meant primarily for online services systems and that properly handles rules heavily dependent of workflow states and business parameters.

We begin by situating the problem from a communication-based view of a workflow system. Next, the metamodel developed is defined and its notable features discussed, and finally, the process of analysis is considered leading to the comparison with other approaches and the conclusions on the adequacy of the process.

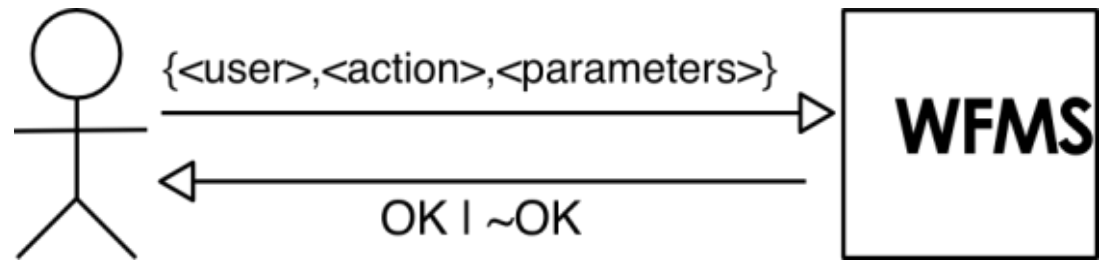

Figure 1. Model of communication with the WFMS.

## II. DESCRIBING WORKFLOWS

The use of Petri nets (PN) [1] for modeling business workflows has been widely accepted for years, mainly thanks to their mathematically sound nature combined with their large power of representation of state-based scenarios [2]. The extended concept of colored Petri nets (CPN) [3] enhances the expressiveness of models and simplifies their analysis, especially when aided by tools such as the CPN Tools software [4]. In [5] these characteristics of PN have been exploited as the authors devised the possibility of linking workflows to multilevel secure environments, thus treating problems of authorization within such workflows as reachability problems in their corresponding modeled Petri net [6]. This approach makes it possible to formally analyze whether a security policy is respected in a given scenario.

Regardless of its structure, a workflow management system (WFMS) may be seen, from its inputs and outputs point of view, as an entity receiving sequences of messages, or *requests*, from the interacting parties that alter its inner state. For the specification of a security policy for such a system, the most important feature is whether or not it authorizes each received request.

Figure 1 indicates the structure of a request. The interacting party that issues the request will be referred to as *user*. Each request specifies a desired *action*, which is subject to security constraints. The set of actions related to the same high-level business process form a *task*. The *parameters* specify the scope of the action, and may be thought of as lists of key-value pairs.

By restricting the mathematical domains for those parts of a request, one may define the set of possible sequences of requests for performing tasks that we shall call *protocol* of the WFMS. Therefore, the description of the sequences of messages that lead to authorized or forbidden actions constitute a formal language, which we'll refer to as the high-level description of the policy in the context of our model.

Based on these formulations, our strategy for validating security policies for WFMS is:

1) Determine the protocol for a given system;
2) Describe the security policy in terms of the authorized and unauthorized sequences of requests;
3) Model the security policy in terms of a special metamodel; and
4) Translate the high-level description of the policy into predicates that query such metamodel's state-space for validating that it fully represents the described security policy.

This paper discusses the design of this metamodel, that must be capable of modeling complex security rules related to business parameters and must also feature an architecture that facilitates the analysis of the models. The analysis suggested in the last step of the strategy defined above is not a complete one. It is enough to demonstrate that models designed in terms of this metamodel properly represent the security policies they were meant to represent, and that the model's architecture supports analyses based on querying its state- space. Additionally, since the metamodel models a complete and consistent set of rules by design, inconsistencies in the rules that guide the query-based analysis will be discovered. However, demonstrating the completeness of this set of rules would require other verifications that although haven't been shown in the scope of this work, also rely on observing aspects of the state-space and can be achieved with no changes to the metamodel.

We focus on verifying that: (a) the metamodel conceived is capable of representing the security policies of the desired scenario without compromising the feasibility of the analysis, and (b) there is a method for translating the typical sequences of messages that will define rules in WFMS into queries that may be designed in the realization of the model using CPN Tools.

### *A. Example: Policy For Online Banking System*

We demonstrate the ideas proposed with the help of an example of security policy meant for regulating the clients' access to an online banking system. The financial services selected for the example are a simple balance check and an electronic funds transfer (EFT) operation. These two, along with the login and logout operations suffice to demonstrate the methodology to be followed and the power and limitations of the modeling.

In this example, each bank account holds two users, one called "master" and another called "helper". Users authenticate via a login procedure where only a password is provided for the sake of simplicity. The balance check is a simple read-only transaction whereas the EFT requires more complex rules that demonstrate how to represent business-specific scenarios.

The protocol for the example is given below (arrows indicate workflow sequence):

Login
Actions: "idtf" (acc, usr) → "auth" (pass)
"idtf" Identifies the account and user for logging in. "auth" sends the password for the (user, account) pair.

Balance
Actions: "balance"
A single message for requesting balance, dependent on the login.

ETF
Actions: "transf_home" → "transf_forms" (acc, val) → "transf_auth" (idt, pass)
"transf_home" represents the request for a funds transfer page containing the required forms. "transf_forms" represents the sending of those forms including account to receive funds and value. "transf_auth" represents the sending of the necessary credentials for confirming the operation.

The rules of the policy are as follows:

1) For a *login*, the requesting user must not have completed a login before under the requested account, unless it has completed a logout in between them.
2) Failing to provide the correct password to the *login* on three consecutive occasions blocks the access to the system.
3) Only logged users may access the account *balance*.
4) Only logged users may access the *electronic funds transfer* operation.
5) Only the "master" user may complete electronic funds transfer operation; the "helper" user may only format them for later approval.
6) The amount to *transfer* to a non-registered account added to the total amount already transferred to non-registered accounts must not exceed the limit of $500.
7) The amount to *transfer* to a registered account added to the total amount already transferred must not exceed the limit of $1500.

Each of these may be specified in terms of authorized and denied sequences of messages, as will be discussed in the analysis of the example. A small CPN implements the sending of sequences of messages to the metamodel, optionally regulating stop criteria.

## III. The Metamodel

The metamodel must provide an abstraction for the inner state of the WFMS as well as include the mechanisms by means of which a modeled policy shall define the logic of authorization of workflows and evolution of the inner state abstraction.

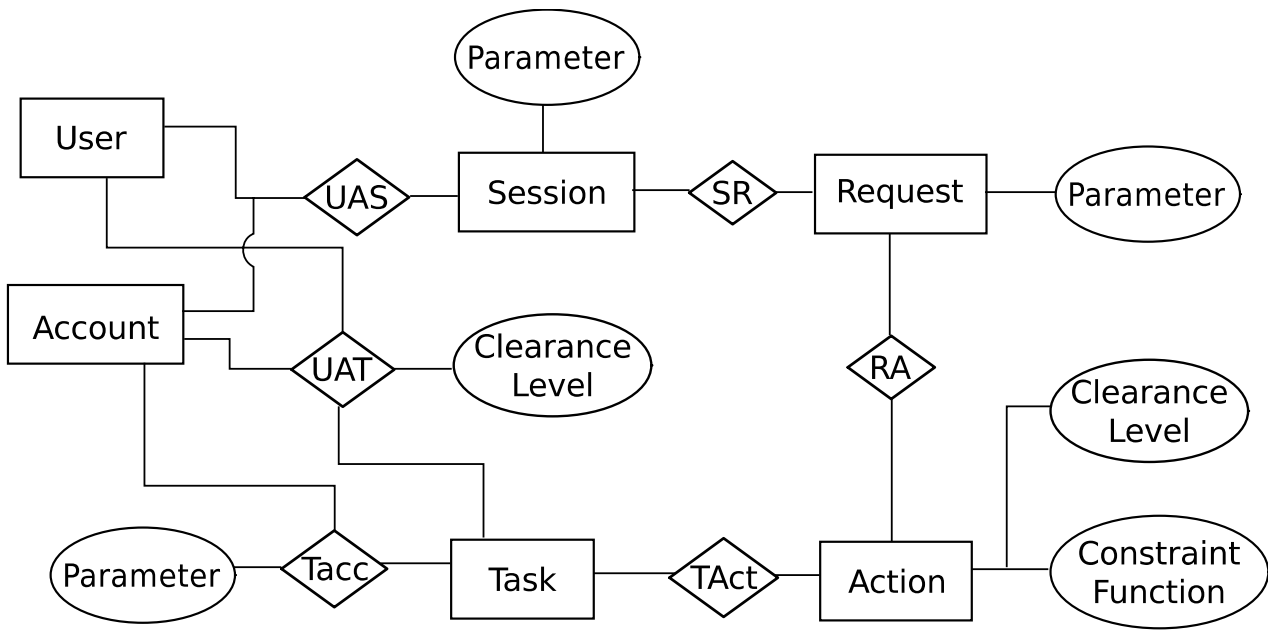

Figure 2. Entity-relationship diagram for the metamodel.

For defining the inner state model, some assumptions on the type of system under discussion are pertinent. Unlike many workflow systems in the literature, online services systems are marked by a wide range of possible operations, or *tasks*, and limited shared resources. In the context of the metamodel, the *account* will act as the only shared resource repository. That'll prove not to be too limiting, since an unlimited number of parameters may be modeled as resources in each account.

Figure 2 presents the metamodel as an entity-relationship model. A *user*, accessing an *account*, initiates a *session*. By means of this session, the requests are issued. Every account holds parameters relative to each performable task. There are also specific parameters for each session and each request. Every triple (user, account, task) is assigned a certain clearance level, and each possible action is associated with a minimum clearance level needed for its authorization. Besides that, every action is also assigned a constraint function that holds the authorization logic for that action in terms of the requesting entities and their parameters. The definitions concerning a security policy modeled on top of this metamodel should, therefore, be achieved by providing values to the depicted attributes — parameters, clearance levels and constraint functions. Taking its formal interpretation as explained in [7], the formal definition of the metamodel follows.

*Definition 1:*

- Let User, Account, Session, Request, Action and Task be entity sets;
- Let UAS $\subseteq$ User $\times$ Account $\times$ Session be a relation assigning a session to a user and account;
- Let UAT $\subseteq$ User $\times$ Account $\times$ Task be a relation assigning a task to a user and account;
- Let SR $\subseteq$ Session $\times$ Request be a relation assigning a request to a session;
- Let RA $\subseteq$ Request $\times$ Action be a relation assigning an action to a request;
- Let Tact $\subseteq$ Task $\times$ Action be a relation assigning an action to a task;
- Let TAcc $\subseteq$ Task $\times$ Account be a relation assigning an account to a task;
- Let Key, Value, Clearance, Integer and String be value sets, with Key $\subseteq$ String, Clearance $\subseteq$ Integer and Value $\subseteq$ Integer $\cup$ String $\cup$ $2^{\text{Integer}} \cup 2^{\text{String}}$;
- Let Parameter $\subseteq$ Key $\times$ Value be a relation assigning a value to a key;
- Let $P_S$: Session $\rightarrow$ $2^{\text{Parameter}}$ be a function that represents a session's *Parameters* attribute by mapping a Session to a set of Parameter (to be defined in the metamodel implementation);
- Likewise, let $P_R$: Request $\rightarrow$ $2^{\text{Parameter}}$ and $P_{TA}$: TAcc $\rightarrow$ $2^{\text{Parameter}}$ be functions representing the analogue attributes; and
- Let $Cl_A$: Action $\rightarrow$ Clearance and $Cl_{UAT}$: UAT $\rightarrow$ Clearance be functions representing the *Clearance Level* attributes;

Given the above definitions, one may formalize the authorization of a request:

*Definition 2:*

- Let $\geq_{Cl}$ $\subseteq$ Clearance $\times$ Clearance be a partial order on the set of Clearances;
- Let $C_A$: Action $\rightarrow$ ($2^{\text{Parameter}} \times 2^{\text{Parameter}} \times 2^{\text{Parameter}} \times$ User $\times$ Account $\rightarrow$ {true, false}) be a higher-order function mapping an Action to a Boolean-valued *constraint function* (referenced ahead as $f_A$);
- Let $\psi$(r): Request $\rightarrow$ {true, false} be an auxiliary predicate such that:

  $\psi(r) := \{ Cl_{UAT}(uat) \geq_{Cl} Cl_A(a) \mid \exists a \in Action : RA(r, a) \wedge \exists s \in Session : SR(s, r) \wedge \exists u \in User, acc \in Account : UAS(u, acc, s) \wedge \exists t \in Task : TAct(t, a) \wedge \exists uat = (u, acc, t) : UAT(uat) \}$

  and

- Let $\varphi$(r): Request $\rightarrow$ {true, false} be a predicate such that:

  $\varphi(r) := \{ C_A(act)\ (P_R(r), P_{TA}(\tau), P_S(s), u, acc) \wedge \psi(r) \mid \exists act \in Action : RA(r, act) \wedge \exists s \in Session : SR(s, r) \wedge \exists u \in User, acc \in Account : UAS(u, acc, s) \wedge \exists t \in Task : TAct(t, act) \wedge \exists \tau = (t, acc) : TAcc(\tau) \}$

Then, a request r is said to be authorized if, and only if, it satisfies the predicate $\varphi$(r).

Predicate $\psi$(r) is the authorization stage that implements multilevel access control by checking whether a certain user working with a certain account has enough clearance for performing its desired action in the context of that specific task.

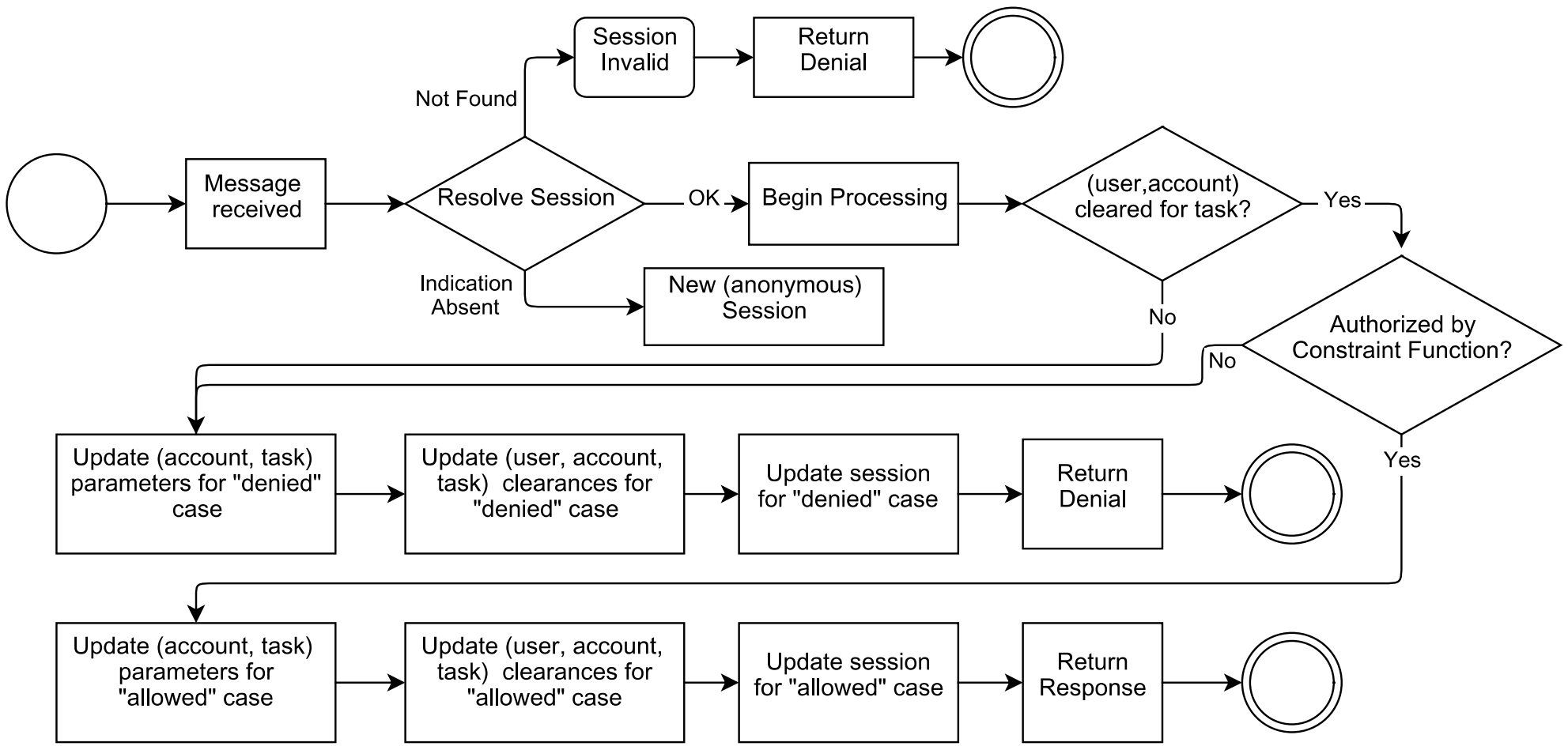

Figure 3. Stages of the processing of a request.

The constraint functions $f_A$, defined for each action in the model, is the placeholder for any complex business logic to be included in a security rule. In a strict view, $f_A$ is capable of including the functionality that ψ(r) provides, but having a separate mechanism for multilevel access control simplifies the modeling given its frequent usage.

Besides the authorization stage, a *state update* stage is desired, as represented by the flowchart in figure 3. By including updates to the model's attributes, decisions regarding the authorization of subsequent requests may differ from previous ones, effectively making the model dynamic. There are three updates contemplated by the metamodel: update of the (account, task) parameters, update of the (user, account, task) clearances and update of the session as a whole (entities, relations and parameters). Adopting a superscript Δ as notation for values during the processing of a request r and Δ + 1 for their updated values during processing of the request that follows r, the update rules are formally defined as follows:

*Definition 3:* Let $A_P$: Task × $2^{\text{Parameter}}$, then we define the *account parameters update function* $f_B$:

$$f_B : 2^{\text{Parameter}} \times 2^{A_P} \times 2^{\text{Parameter}} \times \text{User} \times \text{Account} \times \text{Task} \rightarrow 2^{\text{Parameter}}$$

And so, for the account *acc* processed in Δ and ∀t ∈ Task:

$$P_{TA}^{\Delta+1}(t, acc) = f_B(P_R(r), \bigcup_{\tau \in \text{Task}} (\tau, P_{TA}^{\Delta}(\tau, acc)), P_S^{\Delta}(s), u, acc, t)$$

where r is the request processed in Δ, s ∈ Session : SR(s, r) and u ∈ User : UAS(u, acc, s).

For any other acc′ ∈ Account, ∀t, $P_{TA}^{\Delta+1}(t, acc') = P_{TA}^{\Delta}(t, acc')$.

This means that only parameters relative to the account that issued the request may be updated, however parameters from different tasks than the one processed may also suffer changes, therefore allowing some dependence between tasks in the security policy design.

For the update of (user, account) clearances, we define:

*Definition 4:* Let $UA_C$: Task × Clearance, then we define the *clearances update function* $f_C$:

$$f_C : 2^{\text{Parameter}} \times 2^{\text{Parameter}} \times 2^{\text{Parameter}} \times 2^{UA_C} \times \text{User} \times \text{Account} \times \text{Task} \rightarrow \text{Clearance}$$

and so, for entities related to the processed request in the same terms as in the previous definition, ∀t ∈ Task:

$$Cl_{UAT}^{\Delta+1}(u, acc, t) = f_C(P_R(r), P_{TA}^{\Delta}(t, acc), P_S^{\Delta}(s), \bigcup_{\tau \in Task} (\tau, Cl_{UAT}^{\Delta}(u, acc, \tau)), u, acc, t)$$

For any other acc′ ∈ Account, u′ ∈ User, ∀t

$$Cl_{UAT}^{\Delta+1}(u', acc', t) = Cl_{UAT}^{\Delta}(u', acc', t).$$

Thus, analogously as with $f_B$, a policy may alter the clearances for any task, but only for the pair (user, account) that issued the request. Finally, for updating the session related elements:

*Definition 5:* We define the *session update function* $f_D$:

$$f_D : 2^{\text{Parameter}} \times 2^{\text{Parameter}} \times 2^{\text{Parameter}} \times \text{User} \times \text{Account} \times \text{Session} \rightarrow 2^{\text{Session}} \times 2^{SR} \times 2^{\text{Parameter}}$$

And so, for entities related to the processed request,

$$(\text{Session}^{\Delta+1}, SR^{\Delta+1}, P_S^{\Delta+1}) = f_D(P_R(r), P_{TA}^{\Delta}(t, acc), P_S^{\Delta}(s), u, acc, s).$$

Each of the function kinds $f_B$, $f_C$ and $f_D$ must have a definition in the metamodel implementation for appliance following *authorized* requests and another for appliance following *denied* requests, as denoted in the flowchart.

As a final remark, one may notice that security policies defined according to this metamodel will be inherently *consistent*, since only a single function for each purpose — authorization or state updates — may be defined for each action defined in the protocol; *complete*, and *non-redundant*, since the logic of authorization is bound to all possible interactions with the system, instead of to rules encompassing sets of interactions, which could leave gaps (incompleteness) or overlap (redundancy). However, precisely for not being a rule-oriented design, the modeled policy must be proven equivalent to our high-level specification, which is the purpose of the analysis.

## A. The CPN Model

The translation of the conceptual metamodel conceived into a CPN model is rather straightforward.

Figure 4 shows the network that implements the metamodel. The central transition tagged "execution" handles the entire process seen in the flowchart from figure 3. The input requests are withdrawn from a queue implemented in the place called "server queue" and responses are sent to another queue in the place "server output". Session identifications are generated in the place "new session pool", and the identification number is consumed only in case a new session is processed. The smaller transition, tagged "session invalid" is triggered only in case a certain session identification received is not found in the open sessions pool, located in the place named "open sessions". That mechanism for the invalid sessions is achieved using a lower priority for the firing of that transition. The tokens stored in the place "open sessions" include four pieces of information: the identification of the session, the user associated with it, the account also associated with it and the parameters of the session. The remaining places, "account data access" and "user data access" hold the parameters linked to each account for every task and the (user, account, task) clearance levels, respectively. The guard function for the execution transition ensures that the selected session corresponds to the one referenced by the parameter with key equal to "sess" in the request, when it is present. The code region for the same transition distinguishes new sessions from sessions retrieved from the "open sessions" place and executes the authorization function determining the decision for the request in process. The authorization function is responsible for both stages defined in the flowchart — the first one, $\psi(r)$, referencing the clearance required for the action, and the second one referencing the constraint function for the action. Finally, the output arcs take care of the update of each entity by calling an execution function which references all the right functions defined in the policy, using the result of the authorization function to determine whether to invoke functions for denial or allowance of execution. The output arc leading to the output queue calls a function that assembles the reply message, also for denial or allowance accordingly.

## B. Implementing the Example Policy

As had been explained in the introduction of the metamodel, the security policy is entirely described for the model by establishing the values for all attributes from the entity- relationship model provided. Table I shows a simplified view of that description as it is implemented for the rules above.

TABLE I. SUMMARY OF MODELED POLICY

| Action | Task | Clearance | $f_A$ | $f_B$ | $f_C$ | $f_D$ |
|---|---|---|---|---|---|---|
| idtf | login | 0 | X | | X | X |
| auth | login | 0 | X | X | X | |
| balance | balance | 1 | X | | | |
| transf_home | eft | 1 | X | | | X |
| transf_forms | eft | 1 | X | X | | X |
| transf_auth | eft | 2 | X | X | | |
| logout | logout | 0 | | X | X | X |

The "Clearance" column indicates the clearance level required for the pair (user, account) to perform the action required by the request. All pairs initiate runs of the Petri net with value 0 (zero), gaining higher clearance levels as they log in or perform other actions. The function $f_A$ is the constraint function that determines whether or not the action may be completed. In the table, cells marked with an "X" indicate that the evaluation of a function of the type given by that column is necessary for actions of the type indicated in the row. For functions $f_B$ (account parameters update), $f_C$ (clearances update) and $f_D$ (session update), an empty cell indicates that for that action, an identity function is used for that feature, which means no change is necessary for that entity. The indication of functions for updates in case of denial of intent are omitted to avoid cluttering, but they shall be necessary at most for the same cases as their allowed update counterparts.

It can be noted from table I that, for instance, the "idtf" action requires a constraint function, a clearance update function and a session update function, but doesn't require an account parameters update function. Indeed, the clearance update function is needed because the clearance for all other tasks is reduced below zero, since the session is about to become associated to a user and account that haven't been validated yet. The session update function is needed to indicate which user and account the session is attempting to impersonate. The constraint function is necessary to verify if a free session identifier for allocation, and finally the account parameters shouldn't be updated, since at this point it is yet unknown whether the user actually has access to the account it claims to have (the limitation of the model of only being able to update the account associated to the running session prevents updates on any valid accounts at this point, since there is no account linked to the session until that very update that $f_D$ from "idtf" intends to execute). A sample of an authorization function ($f_A$) for the "transf_auth" action is given below:

```
fun transf_auth_funA (m:request, s:session, q:params) =
let
  val password = StringInParam(getOpt(valueForKey
    ("u"^(toString(usrNumber(#3 s)))) (q),ValString("")))
  val registered = IntListInParam(getOpt(
    valueForKey "registered" (q), ValIntList([])))
  val avLimit = IntInParam(getOpt(
    valueForKey "avLimit" (q), ValInt(50000)))
  val avLimitRegistered = IntInParam(getOpt(
    valueForKey "avLimitReg" (q), ValInt(150000)))
  val tid = IntInParam(getOpt(
    valueForKey "tid"(#3 m), ValInt(~1)))
  val destAcc = AccInParam(getOpt(valueForKey
```

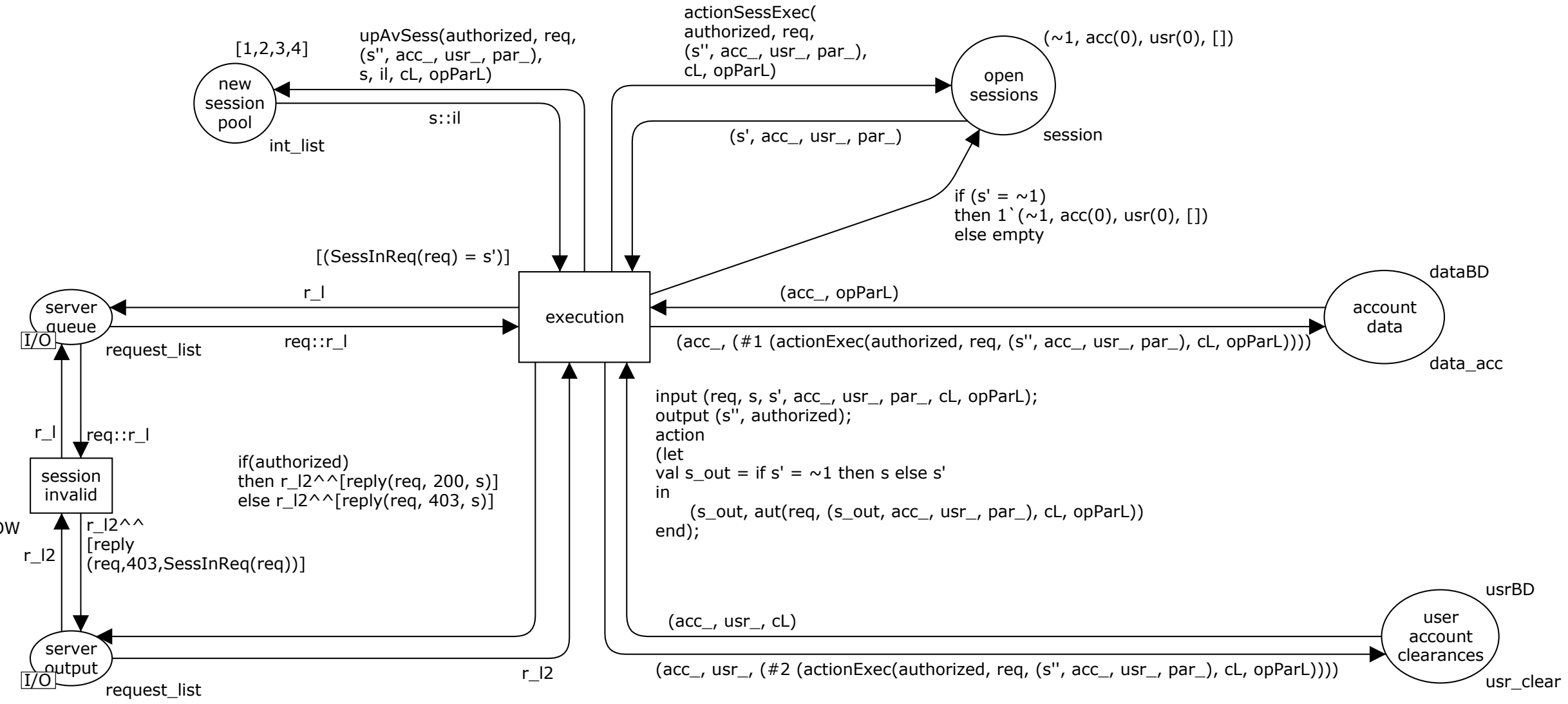

Figure 4. The colored Petri net for the metamodel

```
 (("tr"^(toString(tid)))^"_to") (q), ValAcc(acc(0))))
  val value_ = IntInParam(getOpt(valueForKey
    (("tr"^(toString(tid)))^"_val") (q), ValInt(~1)))
in
  value_ > 0
andalso
  destAcc <> acc(0)
andalso
  destAcc <> (#2 s)
andalso
  if (mem registered (accNumber(destAcc)))
  then avLimitRegistered + avLimit >= value_
  else avLimit >= value_
andalso
  StringInParam(getOpt(valueForKey "auth" (#3 m),
    ValString(""))) = password
andalso
  sessOK(m, s, q)
end;
```

The function's input matches the definition for constraint functions: the value `m` holds the request parameters, `q` holds the parameters for the pair (account, task) and `s` is a triple containing user, account and session parameters. The auxiliary values `password`, `registered`, `avLimit` and soforth are all extracted from the account parameters for the EFT task (`q`) except for `tid` which is a parameter from the request. `tid` identifies the EFT previously prepared for execution, and, therefore, its parameters of value and recipient account are located by referencing it. Default values are specified for all parameters in case they aren't found. The authorization is granted given that the transaction value is larger than zero, the recipient account is valid and isn't the session's own account, the password provided in the request matches the saved password for that user in the account's parameters, and the account's limits are greater than the transaction value, with the appropriate limit calculated depending on whether the recipient account is registered or not. This sample fully demonstrates the complexity that can be achieved in the semantics of the rules thanks to the minimal restrictions provided by the metamodel.

# IV. Model Analysis

## A. Defining Rules Precisely

The main goal of the metamodel analysis as we have conducted it is demonstrating that it fully represents the security policy described by means of accepted and rejected sequences of requests. In order to indicate these sequences and define rules precisely, a notation is introduced. The expression below indicates rule (1) from the example written using such notation:

$(u, a)\text{“auth”}(sess = s)()^A \rightarrow \neg((u, a)\text{“logout”}(sess = s)()^A) \Rightarrow (u, a)\text{“idtf”}(acc = a, usr = u)^D$

$\alpha \Rightarrow R^\delta$ means that if conditions $\alpha$ are respected, decision $\delta$ must be applied to request R. $R^D$ indicates the denial of R. $R^A$ indicates the authorization of R. Operator "$Q \rightarrow R$" indicates R has been processed after Q (not necessarily *immediately* after). Operator $\neg R$ indicates R has not been processed. The first pair of parentheses after each action name encloses request parameters and the second, response parameters (hence, there is no second pair of parentheses for the request under analysis). Thus, the rule reads: "Requests from user *u* for action *idtf* on account *a* shall be denied if there has been a previous authorized processing of an action *auth* for the same user and account in a session *s* that hasn't been followed by an authorized request for action *logout*, also in session *s.*"

It is not our purpose to formalize this notation in this paper. We employ it simply as an intermediate step for designing predicates over state-the space that capture the semantics of the rules written in natural language.

And so, for the remaining rules from the given example of policy:

2) $(u, a)\text{“auth”}()()^D \rightarrow \neg((u, a)\text{“auth”}()()^A) \rightarrow$
$(u, a)\text{“auth”}()()^D \rightarrow \neg((u, a)\text{“auth”}()()^A) \rightarrow$
$(u, a)\text{“auth”}()()^D \Rightarrow (u, a)\text{“_”}()^D$

3) ¬((u, a)“auth”(sess = s)()$^A$ ) $\Rightarrow$
(u, a)“balance”(sess = s)$^D$

4) ¬((u, a)“auth”(sess = s)()$^A$ ) $\Rightarrow$
(u, a)“transf_home”(sess = s)$^D$

5) (u, a)“transf_auth”()$^D$, u = “helper”

6) [(u, b)“transf_forms”(val = v, dest = a)(idt = t)$^A$ $\rightarrow$
(u, b)“transf_auth”(idt = t)$^A$]$^i$ $\rightarrow$
(u, b)“transf_forms”(val = $v_k$, dest = $a_k$)(idt = $t_k$)$^A$ $\Rightarrow$
(u, b)“transf_auth”(idt = $t_k$ )$^D$,

$$a_k \notin R_b \wedge v_k + \sum_i (v_i \mid a_i \notin R_b) > 500 :$$

(“registered”, $R_b$) $\in P_{TA}$(ETF, b)

7) [(u, b)“transf_forms”(val = v, dest = a)(idt = t)$^A$ $\rightarrow$
(u, b)“transf_auth”(idt = t)$^A$]$^i$ $\rightarrow$
(u, b)“transf_forms”(val = $v_k$, dest = $a_k$)(idt = $t_k$)$^A$ $\Rightarrow$
(u, b)“transf_auth”(idt = $t_k$ )$^D$,

$$a_k \in R_b \wedge v_k + \sum_i v_i > 1500 :$$

(“registered”, $R_b$) $\in P_{TA}$(ETF, b)

For properly analyzing the conceived scenario, besides these 7 rules, each restriction on workflow sequence must also generate an additional rule, stating, for instance, that the approval of an EFT (action “transf_auth”) must always follow its definition (action “transf_forms”).

### B. *Writing State-Space Predicates*

Given a set of initial conditions, namely the pre-programmed parameters of each user and account and the set of all (relevant) possible requests for the WFMS, the resulting CPN of the model can be analyzed to generate the graph of all possible states it may assume. Since the number of these states is most likely too large to allow a manual analysis, CPN Tools allows the modeler to automate the search for states with specific properties by executing queries that describe these properties and filter the state-space.

Every security rule described as above states a sufficient though not necessary condition for the outcome of the request to which it refers. Consequently, if a rule states that a request R is to be denied when it satisfies the conditions α, to test that rule one must search for states where R satisfies the conditions α, and yet it has been authorized. Since the state-space analysis is designed to cover all possible situations, if no such state can be found, then the rule has been followed.

In CPN Tools, the function `PredAllNodes` allows one to filter the generated state-space according to some predicate function, usually stating properties of a marking. Also, combining functions `InArcs` and `SourceNodes`, one may obtain a list of the immediate predecessor states to any desired state. Applying these recursively, it is possible to generate the ordered lists of all acyclical paths leading to the states that satisfy the right-hand side of any rule. With the help of functions `PredAllSccs` and `SccToNodes`, it is also possible to determine all sets of states forming cycles in the state-space graph. Therefore, testing any rule expressed in our given notation becomes a matter of verifying the presence or absence of states representing the rule’s restrictions within the lists for some or all paths leading to a set of states.

For most rules, however, some simplification is possible and desirable for optimizing performance. The following pseudocode shows the structure of the query for rule (7), which is a good example of a highly complex rule.

```
auths ← all states where a “transf_auth” action has been
authorized
for all states s1 in auths do
  if s1 is in a state-space cycle then
    add s1 to insecure_states
  end if
  for all states s2 in acyclical paths p from s1 back to 1 do
    if a “transf_auth” action has been authorized in s2 and
    account(s2) = account(s1) and user(s2) = user(s1) then
      add parameter tid from s2 to list t
    end if
    if a “transf_forms” action has been authorized in s2 and
    account(s2) = account(s1) and user(s2) = user(s1) and
    parameter acc from s2 is in the registered accounts list
    from account(s1) and parameter tid from s2 is in list t
    then
      limit_consumed ← limit_consumed + (parameter val
      from s2)
      remove tid from list t
    end if
  end for
  largest_limit_consumed = max(limit_consumed) in p
  if largest_limit_consumed > 1500 then
    add s1 to insecure_states
  end if
end for
return insecure_states
```

The syntax and auxiliary functions for navigating the state-space are all properly documented in [8].

With this logic for rule validation, there is no need for the modeler to tamper with the state definitions adding extra information to function as clues for identifying the trail of states while analyzing a single PN marking. Such a technique will always increase the total number of states in an analysis. Another avoided pitfall is the writing of predicates that reason about inner states of the model linked to decisions about modeling rather than system specification — doing so increases the risk of using false arguments to attest properties of the model.

### C. *Preventing State-Space Explosion*

In many workflow systems, as is the case in [6], [9] among others, the possible sequences of actions that can be requested by a user are few, and may be completely included in the model. However, there are systems where a wider variety of possibilities exist, causing any attempt to model all possible workflows to generate a state-space too large for analysis. For such cases, the analysis must be subdivided in a way that combining each division’s independent analysis yields the same conclusions as the analysis made as a whole.

It is reasonable to assume that, for most systems, different tasks often consist of independent workflows and, therefore, tasks or groups of tasks could be the pivotal elements of the necessary subdivision. Our metamodel, expecting such a need, effectively separates the data domains that constitute the inner state of each task. Let us recall that the decision regarding a request in a WFMS is a function of the parameters from a pair (account, task), whereas the update function for these parameters is executed for all tasks. In practice, this means that when updating a state, a request from a certain task may or may not alter the parameters for another task at will, but the decision is always based solely on parameters exclusive for the task that encompasses the processing request.

By cleverly comparing the parameters of a task in the states immediately preceding and immediately following the processing of a request, we may conclude whether that request causes a change in state in the scope of the task observed. If by doing so for both an authorization and a denial of the same action, we verify that the parameters remain unchanged, than that action is proven independent of the observed task, and may be excluded from the analysis of that particular task. Exceptions must be made for the cases when the action alters the session parameters, which are shared globally by all tasks, and, more subtly, when that action influences the outcome of a different action, which in turn affects the task under analysis.

As an additional simplification, unless the policy contains some rule such as rule (2) from the example, where consequences of a denied request are specified, the update functions for denied cases should not alter the state of the system and, therefore, the receiving of a denied response could be used as condition for terminating a workflow, preventing several cyclical paths from being calculated and speeding up the evaluation of predicates.

Preventing state-space explosion involves making intelligent assumptions or simplifications during the modeling [10] and the acceptable limits of state-space size and calculation time depend on the application. Merely as a reference, table II lists the number of states generated for various conditions in the example policy based on an initial set of 7 well-formed requests, one of each existing actions, requested by a "master" user in account 1.

The results shown indicate that the state-space is generally more sensitive to an increase in the number of different accounts than request variations on its workload. That fact may be understood as the effect of the several different intermediate states caused by the multiple possible orderings in which each request may be sent to the server when many clients are accessing it simultaneously. It is important to notice, though, that tolerating larger workflows for a single client is a very positive feature, since it allows the conception of test cases that test the dependency between sequences of operations in the model, which is the likely case where no subdivision in the suggested fashion is possible. Putting together that fact, the various possible mechanisms discussed for reducing state-space size, the treatability of the general structure of queries for rules, and the successful analysis of the example case, there is good evidence of the adequacy of the modeled policy for the intended analysis.

TABLE II. NUMBER OF STATE-SPACE NODES FOR DIFFERENT WORKLOAD CONDITIONS

| Workload condition (in relation to base case) | Request variations | States |
|---|---|---|
| Base case | 7 | 1122 |
| Wrong login password | 7 | 86 |
| Wrong EFT password | 7 | 470 |
| "helper" user | 7 | 470 |
| EFT of $500 instead of $250 | 7 | 860 |
| + Request for "transf_auth" with other "tid" | 8 | 1845 |
| Two previous tests combined | 8 | 1113 |
| Misc. variations of parameters | 14 | 58911 |
| Base requests also for "helper" on same acc. | 14 | 13997 |
| Base requests also for "master" on other acc. | 14 | 104320 |

## V. RELATED WORK

A significant difference between our approach and all others in the line of [6], is that their analyses [11], [12] are focused on finding a state with insecure properties whereas ours deter- mines an insecure condition by locating an insecure state *path*. By adopting this concept, we introduce a trade-off between state-space graph search time and state-space size, which, to our knowledge, hasn't been investigated in the literature for this area.

The definition of a *value dependency* given in [11] suits our ultimate CPN translation of a complex business rule, however, modeling as they propose requires knowledge of all possible outcomes of calculations at design time making the task impractical whereas, thanks to the concept of colored tokens, we are able to differentiate states assigning the result of a calculation to a token simplifying the design.

Other differences include our definition of a metamodel in a higher level of abstraction, which allowed us to adopt certain general assumptions in the analysis. As a side note, the example model from [12] of a document release process could be modeled using our strategy by representing the document resource as a property in an *editing* task within an account, and setting its value to represent the user currently allowed to perform actions in its workflow.

A different approach, adopted in SecureUML [13], aims at dealing with complex systems security by orienting their design and translating the resulting specification into a formal security policy model. Even though their metamodel is generally more comprehensive than ours, it is not targeted at dealing with workflows and complex business logic. Moreover, the analysis they propose [14] mentions that support for handling system state, which could include the analysis of workflows as we propose, would require reasoning about consequences of their specification's formulas, and has been left for future work.

More recently, a process of analysis of RBAC models [15] in workflows using CPN has been described [16] that shares many characteristics with ours. Much like with the previous examples, this formalization also lacks the ability to express constraints related to the business parameters.

Finally, in [9], the authors define an approach to testing which closely resembles ours, in that the generation of mutants is equivalent to our enumeration of possible input requests. Besides the omission in treatment of business-related rules as in the previous models, the authors mention that for larger systems, analyzing reachability trees could require dividing the system into independent submodules but provide no insight into how such division could be handled. By introducing the notion of restrict data domains for state updates, we have taken a larger step in providing an orientation for the division of these larger systems.

## VI. CONCLUSIONS

The success in modeling the security policy from the example indicates that we have achieved a definition for the metamodel that satisfies the requirement of expressing complex authorization logic linked to parameters from the business model. The communication-based description of rules and its translation into predicates of the model's state-space provide a viable method of ensuring the model's proper behavior and guaranteeing consistency in a given set of rules. The state-space explosion problem was avoided by means of a combination of minimal metamodel design, state-space queries that include conditions on paths to states, and especially a roadmap for subdivision of analysis with guaranteed equivalence of results.

The method of analysis discussed is sufficient to attest whether a modeled security policy is consistent. However, demonstrating its completeness and non-redundancy requires not only the conclusion that the metamodel fully represents the policy's description as also that they are equivalent. A possibility within the existing framework is to analyze the metamodel state-space and derive a set of rules from the behavior it implies, later matching those rules to the original policy description. Since the proposed definition of metamodel also supports that method of analysis, we have achieved our goal of providing an approach to modeling security policies rich with business logic that is suitable for a complete analysis.

We believe that security policy models built with the formalization provided here result in specifications that represent systems behavior in a low level of abstraction, simplifying their implementation in actual code and bringing an extra value to its adoption.

## VII. FUTURE WORK

As outlined above, a method for ensuring completeness and non-redundancy of a policy specification is desired. The method should also include a formalization of the communication-based description language for aiding precise specifications.

Additionally, another dimension of data referring to *context* is desired in the metamodel for signaling overall states of the system, such as "Wednesday" or "raining", to be controlled by special system requests included in the workflow. Other changes allowing a more direct modeling of RBAC and role administration as well as simplifying safe subdivision of analysis are also intended.